# Controllable vs uncontrollable nonlocality:

# Is it possible to achieve superluminal communication?

Fred H. Thaheld

**Abstract**

For over 7 decades it has been debated as to whether nonlocality is strictly of an uncontrollable nature. Based upon polarization measurements with entangled photons, it has been revealed that nonlocality cannot be used to achieve controllable superluminal communication. A simple series of experiments is proposed which may reveal that a rudimentary form of controllable superluminal communication is possible, using human neural stem cells mounted on microelectrode arrays, subject to laser stimulation at varying Hz levels.

**Introduction**

The question of nonlocality was first raised by Einstein-Podolsky-Rosen (EPR), who claimed that if quantum mechanics were a complete model of reality, then nonlocal interactions between particles had to exist (Einstein et al, 1935). Bell later directly addressed this problem of nonlocality, and proposed an experiment to test for its validity (Bell, 1964). An experiment was later performed that showed that nonlocal influences do exist once these particles interact and, that one can test the explicit quantum nature of systems via the use of EPR nonlocality (Aspect, 1982). And, as per Feynman, since this nonlocality cannot be duplicated by a classical system, this enables it to be used to test the quantum nature of systems (Feynman, 1982).

While it has been shown through experiments that superluminal effects do indeed exist between entangled particles such as photons (Aspect et al, 1982; Zbinden et al, 2001), it is generally thought that this effect cannot be used to achieve controllable superluminal communication (Shimony, 1984). Based upon what is known as the Eberhard theorem, it is felt that no information can be transferred via quantum nonlocality (Eberhard, 1978). Relativity theory postulates the non-existence of faster-than-light 'signals' but, does not necessarily impose an analogous requirement upon all other conceivable kinds of 'influences' (Stapp, 1988). He proposes the existence of superluminal influences between the entangled photons which are not to be considered 'signals', with the result that no conflict with the theory of relativity is entailed.

**Analysis of the entangled photons in the Aspect experiment**

Why is it that the entangled photons in the Aspect experiment cannot be used to transfer information faster than light, keeping in mind when commencing this analysis, that we are dealing with *inanimate* entities such as photons, electrons, etc. After all, it has been shown that nonlocal correlations exist between these photons and one would logically think that you should be able to perform this feat. The polarization correlations cannot be used to transmit information faster than light because they can only be detected when the statistics from the measurements on each side are compared in a classical fashion, which is dependent upon the efficiency of the detectors. The act of polarization measurement on photon v1 forces it to move from the quantum indeterminate level to a specific determinate level, and it is this information that is transmitted nonlocally to entangled photon v2. When Alice, let us say, measures the polarization of photon v1 along a direction she chooses, she cannot choose the result nor can she predict what it will be.  Once she makes her measurement, Bob's photon v2 simultaneously receives nonlocal information regarding a similar state of polarization, where he cannot choose the result either.  Since Alice has no control over the results she gets, she cannot send any meaningful information of her own to Bob.  Similarly, Bob can choose one of several polarization measurements to make but, he will not know the result ahead of time.  Alice and Bob can only see the coincidence of their results after comparing them using a conventional method of communication, which does not send information faster-than-light.

**Biological nonlocality: Evidence for entanglement at the *animate* level implying controllable superluminal communication**

I feel that while the above reasoning is correct for entangled *inanimate* or *nonliving entities* such as photons where, once a measurement has been made, the wave function collapses and they become disentangled, the situation is much different for entangled *living entities* such as human subjects and human neurons grown in special basins (Standish et al, 2003; 2004; Wackermann et al, 2003; Pizzi et al, 2004a; 2004b; Thaheld, 2000; 2001; 2003; 2005).  While ground breaking research has been conducted, which indicates in preliminary fashion that entanglement and nonlocality *appear* to exist between human subjects (as per the above references), in this paper I will be concentrating only on the human neurons due to the much better signal-to-noise ratio at mV levels, and the higher degree of replicability.  That some of these techniques (especially regarding the use of anesthetics) may also be applicable to the human subjects, has previously been pointed out (Thaheld, 2005).

In a recent paper (Thaheld, 2000) I had proposed an experiment to determine if EPR-style nonlocal correlations might exist between two neuron transistors, or between neurons grown and mounted on what was then referred to as neurochips. The idea was to stimulate one neuron or a group of neurons, achieve an electrical signal and see if a simultaneous or correlated signal was picked up or elicited by a $2^{nd}$ single neuron or group of neurons, spatially separated and, with both groups in their own individual Faraday cages to rule out most electromagnetic influences of interest, and any acoustical, optical, electrolytic or neural influences. These neurons would have come from the same host, thereby possessing the same DNA genetic characteristics.

A few years later a much more sophisticated and improved version of this proposal was performed by a research group at the University of Milan, utilizing neurons derived from the same human neural stem cells, cultured on microelectrode arrays (MEAs) contained in 2 cm dia basins inside Faraday cages, utilizing low energy lasers to stimulate the neurons (Pizzi et al, 2004a; 2004b). For the purposes of their experiments, they usually have 2 of these neuronal basins separated by 20 cm or more. The voltages in these neuronal basins prior to laser stimulation are ~ 5 mV peak to peak. Laser stimulation of just *one* of the basins at 670 nm, naturally results in an electrical signal being generated by this basin, which can vary from 0-2,000 Hz and, with a peak to peak amplitude of 20 mV. There is a normal delay between the time of activation of the laser, its impingement upon the neurons and the resulting electrical signal, of some 300 ms. What they have found most interesting over a period of several years and thousands of laser pulses, is that the *separated non-stimulated* basin displays simultaneous or correlated electrical signals, of a similar amplitude in mV, and that there is a simultaneity or correlation in the 2 basin's frequencies, between 500-2,000 Hz, with a sharp common peak around 900 Hz. Periodograms of the two signals are also about coincident.

They have resorted to every possible technique (as detailed in their papers) to rule out the possibility that this effect might be due to some error in their experimental protocol, in equipment malfunction or some type of human intervention, since the 2 neuronal basins are separated by only a few cm. To show you the extent they have gone to, to rule out any possible local or classical explanation, they first start out with just one neuronal basin and 2 control basins, without any neurons in them but, containing either their culture liquid or matrigel. Both basins are inside Faraday cages. They stimulate the main basin containing the human neurons on the MEAs with the laser, and get the usual electrical signal response from it but, there are never any electrical signals or response from either of the two control basins.

Then, in other experiments, involving learning in human neural networks on single MEAs (Pizzi et al, 2006), in order to stop these neuron's electrical activity, to insure that the collected signals are actually due to the electrophysiological functionalities of these neurons, the neuronal culture is injected with Tetrodotoxin (TTX), which is a neurotoxin able to abolish action potentials. (As a side note of interest, this is the same neurotoxin which kills several Japanese every year when they eat a popular delicacy known as pufferfish, which has been improperly prepared!). They once again stimulate this TTX treated basin with the laser and get no electrical response, showing the importance of the action potentials, or rather the pores of the voltage-gated sodium channels in nerve cell membranes, since TTX binds to these pores, thereby blocking the action potentials in the nerves. The TTX is then rinsed away, the action potentials return, they stimulate with the laser once again and note that the previously observed electrical signals have returned. This process reveals that the collected signals are actually due to the electrophysiological functionalities of these neurons. As I have previously pointed out, this same technique can also be used with other types of anesthetics for both the neuronal basins and in the case of the human subjects, using anesthetics of a safer and more controllable nature (Thaheld, 2005).

**Proposed experiments to determine if controllable superluminal communication exists**

Based upon the experimental protocol outlined above, it now appears possible to design an experiment which might reveal the existence of controllable superluminal communication in the following fashion.

1. We have 2 separated neuronal basins and stimulate just one of them as usual with the laser, to make sure that the non-stimulated basin still responds with its correlated or auto-correlated electrical signals.

2. We now *reverse* the process and move the laser to the formerly non-stimulated neuronal basin and stimulate it. We want to see if the situation is reversed as regards these two basins. We have in effect exchanged information between them in nonlocal fashion or, by performing a measurement on one neuronal basin, have caused a measurement to take place on the 2$^{nd}$ neuronal basin. Now, if I separate these 2 basins even further and have 2 human subjects with control over their respective lasers, they could send very rudimentary (agreed upon in advance) frequency signals between their 2 neuronal basins. They would not need to compare results after the fact via a classical means, as the variation of the frequency signals would inform each of them if the other party had received the information contained therein. This could take the form of a rudimentary Morse code (Grinberg-Zylberbaum et al, 1994), or a type of binary code consisting of 1s or 0s based upon the manipulation of the laser frequency of stimulation. This can be premised on the fact that neurons have a normal fast firing rate of ~ 1 ms. It is also not impossible that the work of the Italian researchers, as regards learning in human neural networks, could provide an additional approach (Pizzi et al, 2006). And, that this might be further explored from the standpoint of a 'transference of conscious subjective experience' (enabling the transference of large blocks of information without the need for the usual sequential computer binary process), although when I proposed this variation, it was in conjunction with the brains of entangled human subjects (Thaheld, 2003; 2005), and not at the level of just a few neurons.

3. To further make sure that we are truly observing controllable nonlocality and, that what we are observing is not the result of some equipment input or flawed experimental protocol, we inject one of the 2 neuronal basins with the neurotoxin TTX, thereby stopping these neuron's electrical activity. We then stimulate the untreated basin with the laser, and should observe the usual electrical signals being generated by this basin but, with no corresponding correlated electrical signals coming from the TTX treated basin. To double check, we next stimulate the TTX treated basin with the laser and we should not only get no electrical signals from it, but also no electrical signals from the untreated basin! This immediately proves 3 things: First, there are no exterior influences or experimental design flaws which could account for these unusual correlations. Second, that the action potential plays a very important role in this process and, that if it is in any way impaired or stopped, the nonlocality between the neuronal basins also comes to a halt (Thaheld, 2005). Third, that controllable superluminal communication is possible, with or without humans in the picture.

4. Taking this a step further, we now rinse away the TTX from the treated neuronal basin and allow the neuronal activity to naturally commence once again. Once it does, we now stimulate it once again with its laser to see if we get the usual electrical signal response. Now comes the interesting part for, if we do, then we want to see if the other untreated and non-stimulated neuronal basin, simultaneously displays once again the correlated electrical signals. If it does, this would be most unusual indeed, since one would have thought that entanglement between the 2 basins would have been broken by the administration of the neurotoxin. If entanglement was broken, then how could it have been reestablished if the 2 neuronal basins remain separated, i.e., did not come into physical contact again like they were in the beginning? We are left with either the fact that they are both in mutual contact by virtue of the same magnetic field (which can probably be ruled out since the 2 neuronal basins are both inside

Faraday cages), with the same gravitational field or through the vacuum field. As a further cross check, we could once again reverse this process.

**Discussion and speculation**

I am sure that your first comment regarding this proposal has to do with, "Why go to all this trouble, when one can probably determine very simply if there is any entanglement and nonlocality between these neuronal basins, by merely separating them by a few meters or more, while they are both inside Faraday cages, and seeing if this effect still exists". If it does, one could then carry out the proposed experiments, otherwise forget it and move on to another project. It is anticipated that this procedure will be undertaken before the experiments are performed.

Your next objection probably deals with the fact that in the case of the Aspect experiment, after the entangled photon twins were created and flew apart in opposite directions (over a distance of ~ 13 m), they used polarizers which were rapidly changed while the photons were in flight, ruling out local influences at the speed of light, thereby proving the existence of nonlocality (Aspect et al, 1982). In contrast, in our experiment the two MEAs may only be 1-2 m apart initially, so how can we possibly rule out any local influences at the speed of light and prove that we have attained entanglement and controllable nonlocality?

Since Aspect and other similar experiments have already revealed the existence of entanglement and nonlocality, if we are able to prove entanglement in this instance, then nonlocality is automatically proven, regardless of the distance between the neuronal basins.

Another objection to this proposal probably has to do with how this *nonlocal information* which is generated by the neurons along with the action potentials following laser stimulation, can possibly cause another neuron or neurons in a non-stimulated basin to fire, since it is supposedly *nothing* or of an *immaterial* nature, possessing no mass or energy. It does it in the same way and in the same fashion as it does between the entangled photons. It will help in the following analysis if you visualize that the entangled photons = the entangled neuronal basins and, that the polarizers = the neurons.

Why is it that it takes photons possessing energy in a laser pulse to cause the neurons in a stimulated basin to fire, and yet just *nonlocal information*, supposedly possessing no mass or energy, is able to cause the neurons in a non-stimulated basin to fire? It appears to be a dualistic process, in that when a neuron fires, we get not only a classical action potential but a nonlocal quantum event or *nonlocal information*, which event is able to trigger a neuron or neurons in a non-stimulated basin, resulting once again in a classical action potential, without violating the conservation laws.

I had addressed this issue in a previous paper (Thaheld, 2005), by asking the question 'what and where is the biological equivalent in a neuron of the polarizer used in physics, and can this be considered as a two-way transducer'? I.e., since the polarizer is able to convert quantum to classical and vice versa, the neurons should be able to accomplish this same feat, converting the quantum or mental to classical or brain, and vice versa.

Several possible locations have been postulated in the neurons, which could be the targets for these *immaterial* or *nonlocal information* mental events. In this regard it has been postulated (Beck, Eccles, 1992; Eccles, 1994), that a non-material mental event (which I am saying is equivalent to a nonlocal polarization event), could influence the subtle probabilistic operations of synaptic boutons (involved in the action potential), focusing attention on the effective structure of each bouton, which are the *paracrystalline* presynaptic vesicular grids, as the targets for non-material events. This means that we may now be able to examine exactly, from an empirical standpoint, where and how the transition is made from a mental or quantum state, to a brain or classical state and vice versa.

When a measurement is made by a polarizer on a photon, this polarization *nonlocal information* is instantaneously sent to and received by the other entangled photon *nonlocally*, thereby changing its state of polarization. This very basic *information* can be looked upon as if it were quantized, one or more quanta of *information* representing vertical polarization and one or more quanta representing horizontal polarization, with neither quanta representing *actual* mass or energy, but able to act upon mass or energy. This same analogy applies to spin up or down, spin right or left, when entangled electrons pass through a Stern-Gerlach device and are measured. The photons, being *inanimate*, cannot regenerate or reestablish their previous entangled state, so one is limited to just one measurement. And, as has already been pointed out earlier, Alice and Bob cannot exchange any useful information or choose what to send to each other in advance of any measurement.

The situation changes with the entangled neuronal basins. It appears that after every measurement, which is initiated by the laser pulse(s), when the laser stimulated basin generates an electrical signal or signals, there is simultaneously generated by this basin one or more quanta of *nonlocal information*, which is instantaneously received by the 2$^{nd}$ non-stimulated basin, as reflected in the correlated or auto-correlated electrical signals. These basins are then equivalent to the photons, in that they are able to send and receive *nonlocal information*. However, the major difference is that the neurons in the basins being *animate*, appear to be capable of maintaining or regenerating entanglement after each laser measurement (as has already been demonstrated thousands of times), or else there would be no correlated electrical signals arising simultaneously from the non-stimulated basins after each cycle of laser stimulation. Nor would the stimulated basin be able to continuously generate and transmit *nonlocal information* after repeated cycles of laser stimulation. And, since there are numerous entangled neurons with their constituent particles in each basin (with special emphasis on microtubules), the amount of quantized *nonlocal information* is many orders of magnitude greater than that involved in polarization measurements.

Furthermore, Alice and Bob are free to *choose in advance* what type of information they want to send to each other, send it in *nonlocal* fashion and know, without any classical means, the content of what was either sent or received.

And, since this *nonlocal information* consists of neither mass nor energy but, is capable of *acting upon* mass or energy, one could philosophically say that Einstein was correct in one of his major pronouncements on this subject, if we merely change where he placed his emphasis. He said that, "*Nothing* can go faster than the speed of light". This can now become, "Nothing *can* go faster than the speed of light"! Approached in this fashion, special relativity is not violated since *nothing* is *something* in the quantum world and really *nothing* in the classical world, yet it can affect *something* in the classical world through some type of quantum-classical transduction interface.

What I am leading up to, and what I have touched upon in the past, is that there now appears to be an *equivalence* of quantum nonlocality and what I have termed *biological nonlocality* and, that they are either two different versions of the same thing or they are the same thing (Thaheld, 2003).

Finally dear readers, I am sure that you may be puzzled about one last thing just as I am. If these two neuronal basins are truly entangled and acting in a nonlocal fashion, if this is occurring at a macroscopic level, what is the role of gravity in this whole affair? Has it become quantized in a very unforeseen and unexpected fashion?